\documentclass[epjc3]{svjour3}
\journalname{Eur. Phys. J. C}
\RequirePackage{graphicx}
\usepackage[utf8]{inputenc}
\usepackage{hyperref}
\usepackage{subcaption}
\usepackage{upgreek}
\usepackage{siunitx}
\usepackage{booktabs}
\usepackage[normalem]{ulem}
\usepackage{mhchem} 
\usepackage[final]{changes}
\usepackage{lineno}
\usepackage[right = 2cm]{geometry}

\graphicspath{{figures/}}

\newcommand{\ChW}[1]{\textcolor{black}{#1}} 

\newcommand{\AP}[1]{\textcolor{black}{#1}} 
\DeclareSIUnit\bar{bar}

\begin{document}

\title{An active Transverse Energy Filter based on microstructured Si-PIN diodes with an angular-selective detection efficiency}

\titlerunning{\parbox{9cm}{An active transverse energy filter}} 
\authorrunning{S.~Schneidewind, \dots, C.~Weinheimer}
\institute{%
Institute for Nuclear Physics, University of Münster, Wilhelm-Klemm-Str. 9, 48149 M\"{u}nster, Germany\label{a}
\and CeNTech and Physics Institute, Heisenbergstr. 11, University of Münster, Germany\label{b}
\and Kirchhoff-Institute for Physics, University of Heidelberg, Im Neuenheimer Feld 227, 69120 Heidelberg, Germany\label{c}
\and \added{Istituto Nazionale di Fisica Nucleare (INFN) – Sezione di Milano-Bicocca, Piazza della Scienza 3, 20126 Milano, Italy}\label{d}
}
\thankstext{email1}{e-mail: sonja.schneidewind@mib.infn.it}
\thankstext{email2}{e-mail: hannen@uni-muenster.de}
\thankstext{email3}{e-mail: weinheimer@uni-muenster.de}
\author{%
S.~Schneidewind\thanksref{a,d,email1}
\and K.~Gauda\thanksref{a}
\and K.~Blümer\thanksref{a}
\and D.~Bonaventura\thanksref{a}
\and C.~Gönner\thanksref{a}
\and V.~Hannen\thanksref{a,email2}
\and H.-W.~Ortjohann\thanksref{a}
\and W.~Pernice\thanksref{b,c}
\and L.~Pöllitsch\thanksref{a}
\and R.W.J.~Salomon\thanksref{a}
\and M.~Stappers\thanksref{b}
\and S.~Wein\thanksref{a}
\and C.~Weinheimer\thanksref{a,email3}
}

\date{\today}
\maketitle
\begin{abstract}
Si-PIN detectors can be microstructured to achieve angular-selective particle detection capabilities, which we call \emph{active Transverse Energy Filter} (aTEF). The microstructuring consists of a honeycomb structure of deep hexagonally-shaped holes with active silicon side walls, while the bottom of the holes is made insensitive to ionizing radiation. The motivation for this kind of detector arises from the need to distinguish background electrons from signal electrons \replaced{solely by their angle of impact}{in a spectrometer of MAC-E filter type}.
We have demonstrated the angular-dependent detection efficiency of self-fabricated aTEF prototypes in a test setup using an angular-selective photoelectron source to illuminate the detector \textcolor{black}{from} various incidence angles.
\end{abstract}

\section{Introduction}
\label{sec:intro}
\begin{figure}
 \centering 
 \includegraphics[width=0.6\textwidth]{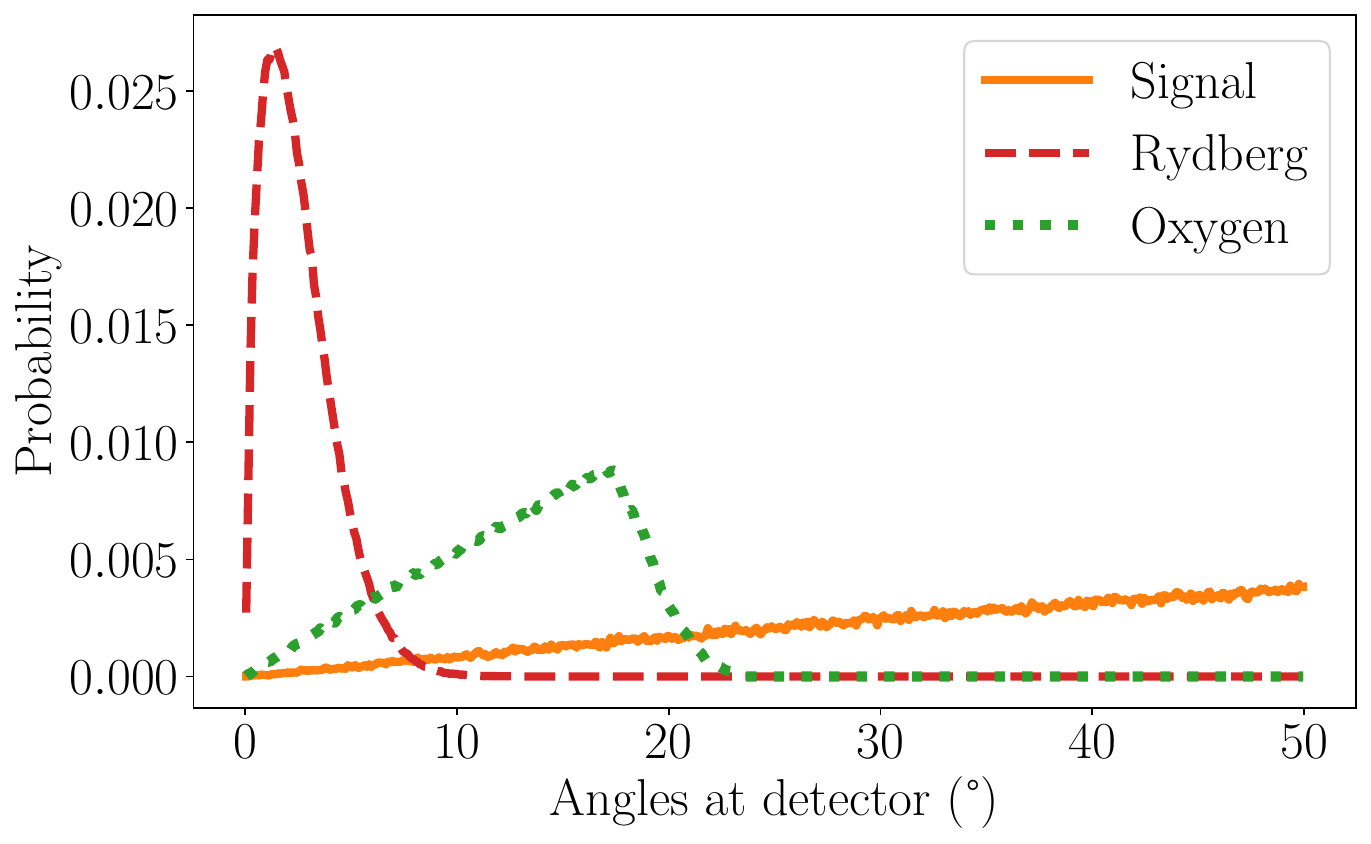}
 \caption{Angular distributions of \added{$\upbeta$-decay} signal electrons (orange) \added{in comparison to} ``Rydberg'' background electrons (red) and background electrons from autoionizing \added{oxygen} states (green) at the detector of the KATRIN experiment, simulated using input from~\cite{PhD-Hinz} and~\cite{PhD-Trost}. The simulation considers the energy spectrum of background electrons in highly excited Rydberg states \deleted{of hydrogen (36\,\%) and oxygen atoms (64\,\%),} ionized by blackbody radiation at room temperature. These Rydberg atoms are sputtered from the spectrometer walls due to $\upalpha$ decays of implanted $^{210}$Pb progenies from the $^{222}$Rn decay chain~\cite{Fraenkle_2022}.
 The Rydberg atom's energies from \cite{PhD-Trost} are corrected for the surface binding energy. \textcolor{black}{After ionization,} the electron energy and direction are calculated with the Doppler shift of the \deleted{Rydberg} atoms taken into account.}
 \label{fig:background_at_KATRIN} 
\end{figure}

Background suppression is a major challenge in experiments with weak signals, such as rare-event searches. While in many cases, signal events in detectors can be distinguished from background events by different energy deposits or different event topology \cite{Wermes}, there are also conditions in which signal and background events are caused by the same type of particle, arriving with very similar energy and differing only by their angle of incidence at the detector. Common angle-sensitive methods, such as low-density tracking detectors or $\Delta E$-$E$ arrangements are, however, not suitable for very low particle energies or in strong magnetic fields with the particles of interest undergoing cyclotron motion.\\
\replaced{An example for such a rare-event search experiment with different angular distributions of background and signal events is the Karlsruhe Tritium Neutrino experiment KATRIN~\cite{ref:TDR2}.}{The Karlsruhe Tritium Neutrino experiment KATRIN~\cite{ref:TDR2} has these physics signal and background features.} Its objective is to determine the neutrino mass from a very precise measurement of the electron energy spectrum of molecular tritium $\upbeta$ decay near its endpoint of 18.57\,keV. It uses a high-luminosity gaseous molecular tritium source with an electrostatic spectrometer with magnetic adiabatic collimation (MAC-E-filter~\cite{mac-e}), which allows electrons above an adjustable threshold with $\cal{O}$(1\,eV) width to be transmitted to a pixelated rear-illuminated Si-PIN detector of 9\,cm diameter~\cite{FPD} \added{and counted there}. The signal electrons gyrate around the guiding magnetic-field lines and arrive with incidence angles up to \SI{49}{\degree} to the surface normal and cyclotron radii $<\SI{230}{\micro\meter}$ at the detector\deleted{  (excluding post-acceleration of $\SI{10}{\kilo\electronvolt}$)}, where the magnetic field has a strength of \SI{2.4}{\tesla}\footnote{\added{For the sake of simplicity of the description the fact, that the electrons at the KATRIN experiment are post-accelerated gaining $\SI{10}{\kilo\electronvolt}$ in kinetic energy before arriving at the detector \cite{ref:TDR2}, is neglected.}}. \\
Most background electrons in the KATRIN spectrometer stem from highly single-excited Rydberg atoms\,\cite{Gallagher_1994} or double-excited\added{, autoionizing} atoms, which are sputtered off from the vessel surface into the spectrometer volume by the recoil of $\upalpha$ decays in the vessel walls\,\cite{PhD-Trost,Fraenkle_2022,ref:SAP,PhD-Hinz}. Ionization of these atoms within the spectrometer volume leads to low-energetic secondary electrons, which are accelerated by the electric potential gradient of the spectrometer and arrive at the detector with similar kinetic energy as the signal electrons. Since the electrons gyrate around the magnetic field lines and accumulate energy parallel to the magnetic field lines by the acceleration, their small amount of energy obtained in the ionization process makes them arrive with a comparably narrow angular distribution relative to the detector's surface normal, see fig. \ref{fig:background_at_KATRIN}. 
Although KATRIN's background has been reduced \replaced{already significantly}{,by a factor of two after restricting the volume of the magnetic flux tube between the highest potential and detector}~\cite{ref:SAP}, the background level in KATRIN is still an order of magnitude higher than anticipated. \deleted{A significant fraction of this background is assumed to be created by ionized Rydberg atoms and autoionizing states.} In reference~\cite{Gauda_2022}, the possibility for background reduction \added{at the KATRIN experiment making use of the different angular distributions} via an active Transverse Energy Filter (aTEF) based on a microchannel plate design was introduced. \deleted{Here,} Motivated by successful etching results through silicon detectors for connecting front and back sides \cite{Garcia-Sciveres:2017ymt}, \added{we investigated} an alternative version \added{of such an angular-differentiating detection concept} using microstructured Si-PIN diodes, the silicon aTEF (Si-aTEF)\deleted{is discussed}. \added{Since microstructuring silicon PIN diodes could be very interesting for other applications, in particular for low-energy rare-event searches, we describe our idea and the corresponding investigations in this paper.} 
The following section provides an introduction to the idea of a microstructured Si-PIN diode as angular-selective electron detector. Si-aTEF prototypes were fabricated \textcolor{black}{at} the Münster Nanofabrication Facility (MNF)\footnote{Münster Nanofabrication Facility, University of Münster, Busso-Peus-Str.10, 48149 Münster, Germany}. \textcolor{black}{T}he applied procedure  \textcolor{black}{of the fabrication of Si-aTEFs} is introduced in sec. \ref{sec:fabrication}. Sec.~\ref{sec:measurements} presents proof-of-principle measurements with \textcolor{black}{the} aTEF prototypes at a test stand in Münster. \textcolor{black}{In contrast to typical detector fabrication processes, a photoresist required for microstructuring is kept on the test samples during the measurements.} Test results regarding a possible performance improvement via \textcolor{black}{removal of the resist combined with} an additional surface-passivation step are discussed in sec. \ref{sec:passivation}, followed by a conclusion and outlook in sec. \ref{sec:summary}.
\section{Idea of \textcolor{black}{a Si-aTEF}}
\label{sec:idea}
\begin{figure}
    \centering
    \includegraphics[width=0.3\linewidth]{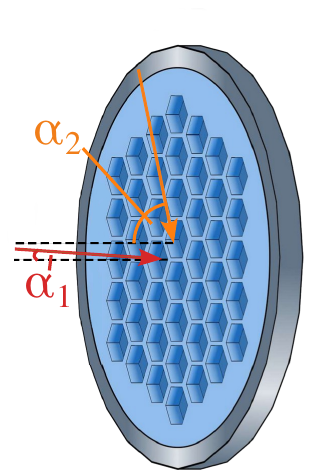}
    \caption{The Si-aTEF is a microstructured Si-PIN diode, in which the individual holes feature sensitive sidewalls and insensitive \replaced{floors}{grounds}. The detector's surface normal is shown as dashed line. Electrons with high incidence angle $\alpha_\text{2}$ relative to the surface normal (orange arrow) have a higher probability to hit a sidewall and thereby induce a
detectable signal, while electrons with low incidence angle $\alpha_\text{1}$ (red arrow) are likely to be absorbed without signal creation.}
    \label{fig:atef-schematic}
\end{figure}
\begin{figure}
    \centering
    \includegraphics[width = 0.7\textwidth]{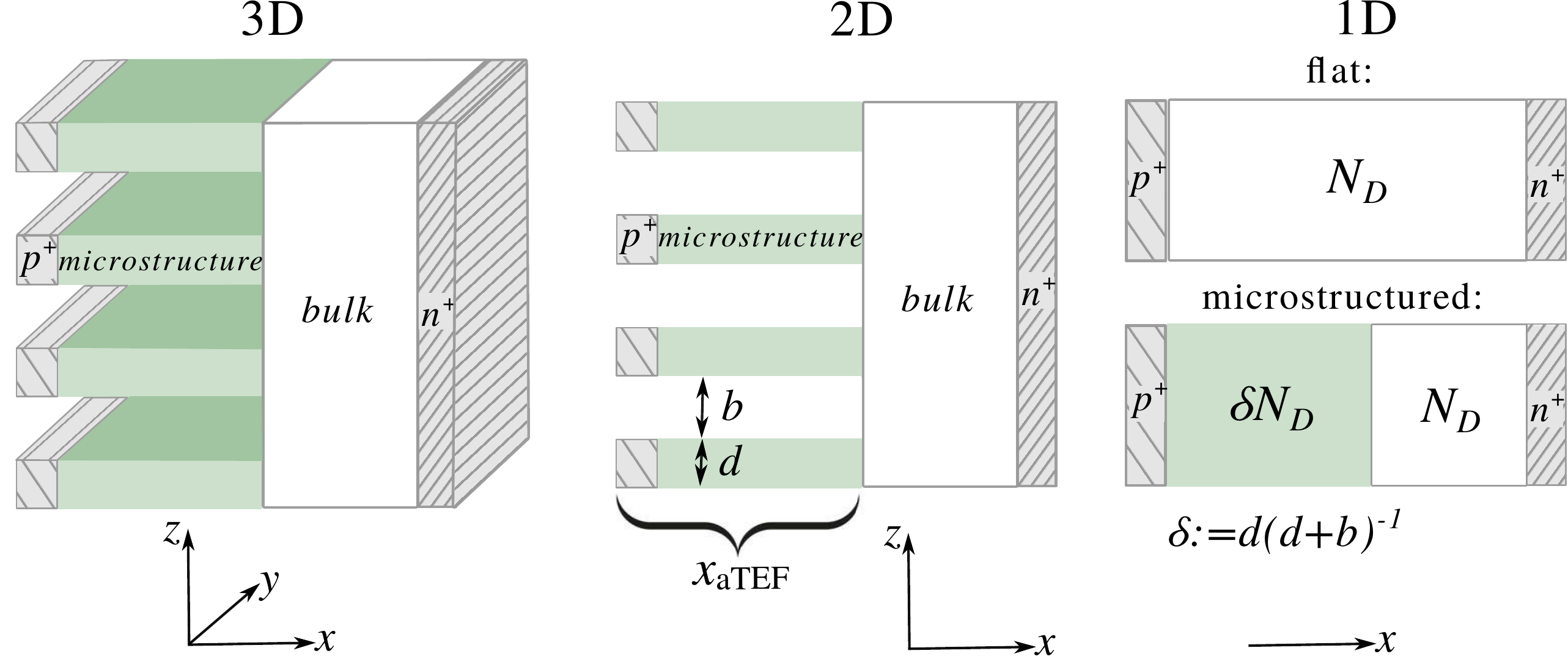}
     \includegraphics[width=0.6\linewidth]{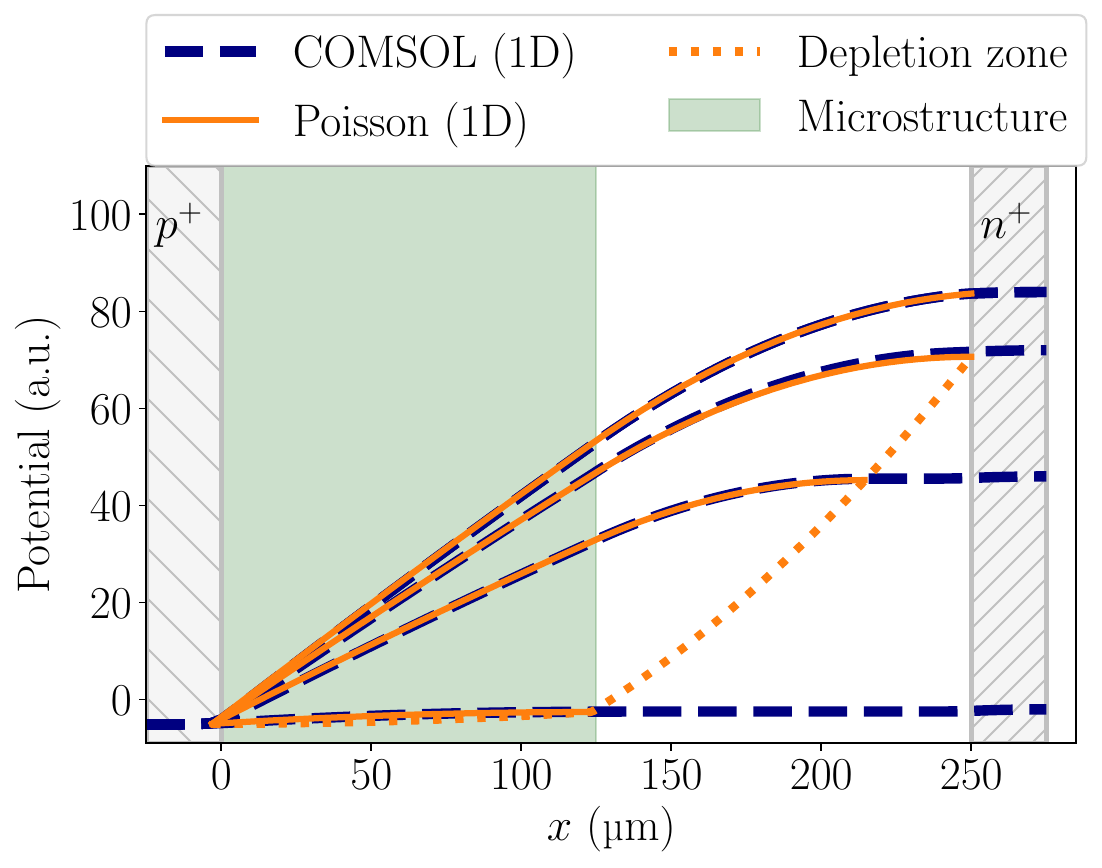}
    \caption{Top: Illustration of the simplification of the 3D-microstructured diode, here for the simplified example with microstructured strips in y-direction instead of hexagons, to a 2D case or two 1D cases with different charge densities. Bottom: Electric potential and width of depletion zone in a hypothetical 1-dimensional frontside-illuminated Si-PIN diode featuring a microstructure with $x_\text{aTEF} = \SI{150}{\micro\meter}$ and $\delta = 0.1$. The \replaced{solid}{straight} orange lines show the curves calculated using the Poisson equation for different $U_\text{bias}$, indicated by the maximum achieved potential. The blue-dashed lines show the same curves simulated using \texttt{COMSOL}. The orange-dotted line shows the potential at which a certain depletion-layer depth is reached. Overall good agreement between calculation and simulation for the 1-dimensional case is found.}
    \label{fig:comsol-simulation}
\end{figure}

The Si-aTEF is based on a Si-PIN diode, as is the KATRIN detector, which is microstructured by hexagon-shaped holes on the illuminated side, as illustrated in fig. \ref{fig:atef-schematic}. The sensitive volume of a PIN diode is the depletion zone with a high electric field gradient. If most of the depletion zone is removed during the microstructuring of the diode with the remaining active material in the channel walls only, low-angular electrons hit the insensitive channel \replaced{floors}{grounds} and are absorbed, while electrons with higher incident angles hit the channel walls and are detected. In this work, \added{all investigations refer to the detection of electrons, but the concept might be applicable also to other ionizing particles or radiation. In this case,} the microstructuring is realized by etching commercially available flat Si-PIN diodes (Hamamatsu\,\footnote{HAMAMATSU PHOTONICS Europe GmbH, Arzbergerstr. 10, 82211 Herrsching, Germany} S3590 photodiodes) via a highly directional silicon-etching process, see sec. \ref{sec:fabrication}. Signal readout is performed in the same manner as for a common diode, using a capacitively coupled charge-sensitive preamplifier. \\
In a microstructured PIN diode, the potential when applying a bias voltage evolves differently from that of a standard one. This is a consequence of the charge neutrality condition, which needs to be adjusted for the microstructured part. The different behavior can be derived in a simplified manner by solving the one-dimensional Poisson equation, as will be shown below. The following considerations are restricted to front-illuminated PIN diodes with $p^+nn^+$-layer structure\,\footnote{As the KATRIN detector is back-illuminated, it would behave differently, see \cite{PhD-Gauda}}, like the photodiodes used in this work. \\
The charge density $\rho$ in the x-direction along the $p$-$n$-transition at $x=0$, assuming a constant doping concentration in the $p$ and $n$ regions, is given by
\begin{equation}\label{eq:chargedensity}
	\rho(x) =
	\begin{cases}
		-e N_A,& \quad\mathrm{for}\quad	 -x_p \leq x \leq 0 \\
		e N_D,& \quad\mathrm{for}\quad 0\leq x \leq x_n
	\end{cases}
\end{equation}
with $x_p$ or $x_n$ being the extension of the space-charge zone into the $p$ or $n$ doped region of the semiconductor. All dopants are assumed to be ionized, i.e., $N_\mathrm{A}^- = N_\mathrm{A}$, $N_\mathrm{D}^+ = N_\mathrm{D}$ for the acceptors ($N_\mathrm{A}$) and donors ($N_\mathrm{D}$), with typical doping concentrations of $N_\mathrm{A} = 10^{19}\,\mathrm{cm}^{-3}$ and $N_\mathrm{D} = 2.3\cdot10^{12}\,\mathrm{cm}^{-3}$ \cite{Wermes}.
Charge conservation in a flat diode implies
\begin{equation}\label{eq:chargeconservation}
	N_\mathrm{A}^- x_p = N_\mathrm{D}^+ x_n.
\end{equation}
Due to the high doping concentration of the $p$-layer, the depletion zone mainly spreads into the $n$-layer.
The width $w$ of the depletion zone further increases when a reverse voltage $U_\text{bias}$, i.e., a positive voltage at the $n$-doped side, is applied. It is given by \cite{Wermes}
\begin{equation}\label{eq:depletionzone_biasvoltage}
	w \approx x_n \approx \sqrt{\frac{2\varepsilon}{e} (U_\text{bias} + U_\text{bi}) \frac{1}{N_\mathrm{D}}}\,,
\end{equation}
with $U_\text{bi}$ \replaced{corresponding}{corresponds} to the barrier potential at the $p$-$n$-junction and the permittivity $\varepsilon = \varepsilon_\text{0}\varepsilon_\text{r}$\textcolor{black}{, with vacuum permittivity $\varepsilon_\text{0}$ and relative permittivity $\varepsilon_\text{r}$}. 
In case of the aTEF detectors the microstructuring and its depth $x_\text{aTEF}$ need to be taken into account.
Fig. \ref{fig:comsol-simulation} (top) shows the assumed \ChW{simplified} diode geometries used to investigate the principal influence of the microstructuring. The 
\ChW{calculation} is simplified by integrating along the z- and y-coordinates, leading to an \ChW{additional modification of the} x-dependent charge-carrier density \ChW{(assuming the $p$-$n$-junction is within the microstructure):
\begin{equation}\label{eq:chargedensity2}
\rho(x) =
	\begin{cases}
		-e N_A \cdot \delta,& \quad\mathrm{for}\quad	 -x_p \leq x \leq 0 \\
		e N_D \cdot \delta ,& \quad\mathrm{for}\quad 0< x \leq x_\text{aTEF}\\
        e N_D,& \quad\mathrm{for}\quad x_\text{aTEF} < x \leq x_n\,.\\
	\end{cases}
\end{equation}
$\delta$ is a reduction factor accounting for the reduced number of available charge carriers in the microstructured region, \AP{which} depends on \AP{the} channel thickness $d$ and channel distance $b$ via $\delta := d(d+b)^{-1}$.}

\begin{figure}
    \centering
    \includegraphics[width=0.6\linewidth]{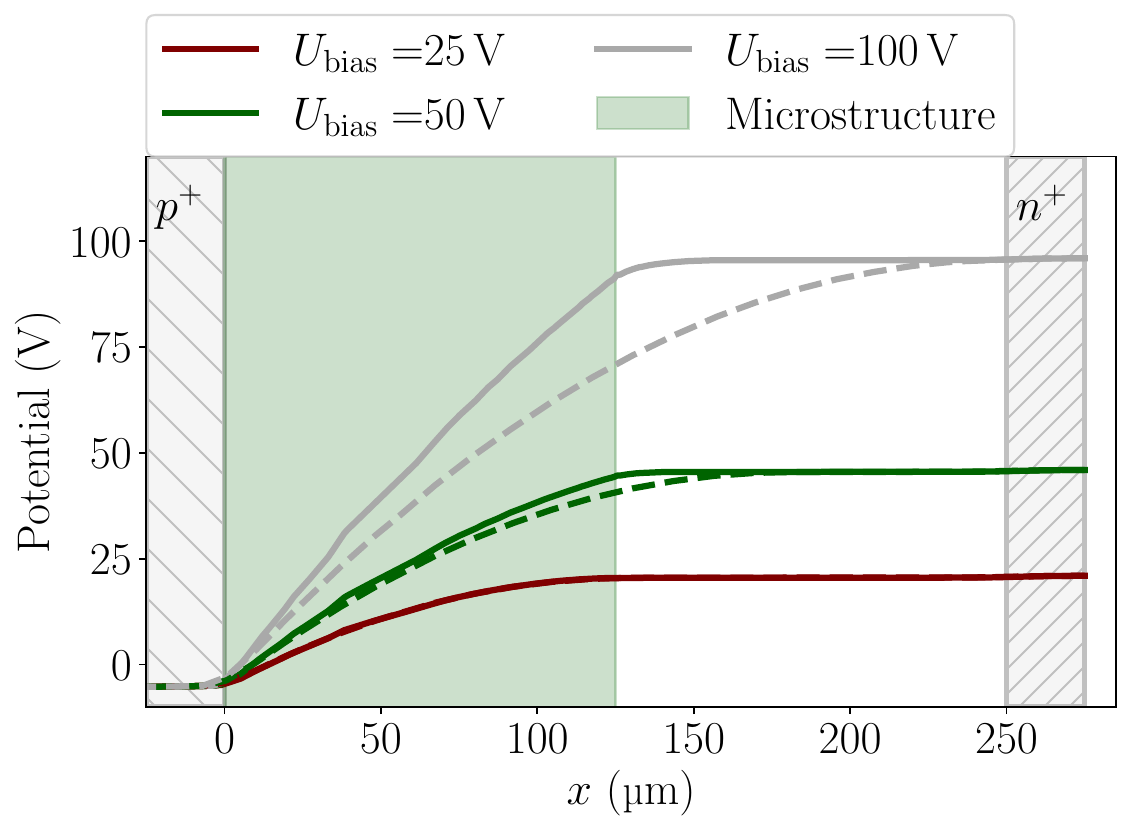}
    \includegraphics[width = 0.2\linewidth]{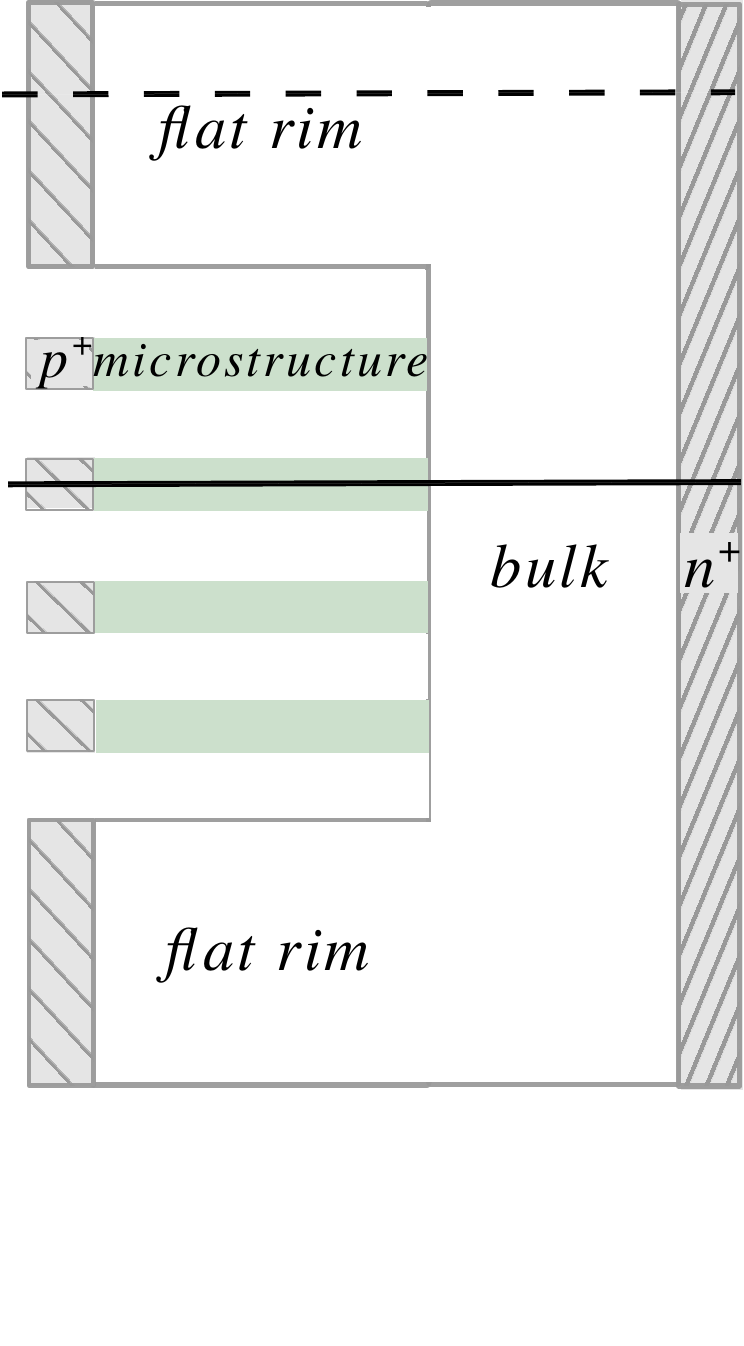}
    \caption{Simulated electric potential profile for a 2D Si-PIN diode which is microstructured in the center and flat at the rim, as illustrated on the right. The \replaced{solid}{straight} lines show the profiles along a channel in the microstructure, while the dashed lines indicate the profiles in the flat rim, which basically behaves as a standard PIN diode.}
    \label{fig:comsol-simulation-voltages}
\end{figure}

This discontinuity at $x = x_\text{aTEF}$ leads to the following electric potential $U$ in the depletion zone (see derivation based on the Poisson equation $\Delta \Phi = \frac{\rho}{\varepsilon}$ with the field $\Phi$ in \ref{appendix}):
\begin{equation}
	\begin{split}
		U(x) &= \int_{-\infty}^x E(x') \,\mathrm{d}x' = \int_{-\infty}^x  \int_{-\infty}^{x'} \frac{1}{\epsilon} \rho(x'') \,\mathrm{d}x''\,\mathrm{d}x' \\
		&=	\frac{eN_D}{\varepsilon}
		\begin{cases} 
			\frac{x^2}{2} - xw &\mathrm{for}\quad U < U_\mathrm{aTEF}\,,\\\\
			x_\mathrm{aTEF} \left( \frac{x_\mathrm{aTEF}}{2} - w_1 \right) - \frac{x_\mathrm{aTEF}}{\delta} \left( \frac{x_\mathrm{aTEF}}{2} - w \right) + \frac{x}{\delta} \left( \frac{x}{2} - w \right) &\mathrm{for}\quad U > U_\mathrm{aTEF}\,,
		\end{cases}
	\end{split}
\end{equation}
with $U_\text{aTEF} := U_\text{bias}(w=x_\text{aTEF})$ and $w_1 = x_\mathrm{aTEF} + \delta^{-1} \cdot (w - x_\mathrm{aTEF})$\,\footnote{Calculation example: $U_\text{aTEF}\approx\SI{30}{\volt}$ for $x_\text{aTEF}=\SI{150}{\micro\meter}$, $N_\text{D} = \SI{2.3e12}{\per cm^{-3}}$ and $\epsilon_\text{r} = 11.68$.}.
Fig. \ref{fig:comsol-simulation} bottom shows the electric potential and the size of the depletion zone for a hypothetical 1-dimensional Si-PIN diode. In addition to the curves calculated using the Poisson equation, comparative field simulations with \texttt{COMSOL}\,\footnote{Comsol Multiphysics GmbH, Robert-Gernhardt-Platz 1, 37073 Göttingen, Germany} Multiphysics are shown, which nicely \replaced{match}{matches} the calculated potential curves.\\
For the 2D case (middle scheme in fig. \ref{fig:comsol-simulation}), an analytical solution of the Poisson equation becomes more challenging. The simulated potential depth profile for \AP{the} 2D case is shown in fig. \ref{fig:comsol-simulation-voltages}, compared for the flat rim and the microstructured center of the diode. Two expectations result from the simulation: First, the microstructure can be fully depleted at \AP{a} comparably low $U_\text{bias}$, leading to sensitive channel walls. Second, the bulk region, including the channel \replaced{floors}{grounds}, depletes only at much higher $U_\text{bias}$ and remains insensitive in most cases.
\section{Fabrication of microstructured Si-PIN diodes}
\label{sec:fabrication}
\begin{figure}
    \centering
    \includegraphics[width = 0.6\textwidth]{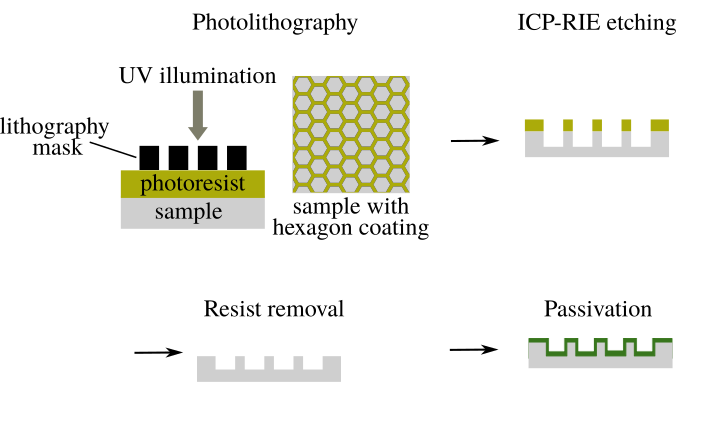}
    \caption{Fabrication steps for microstructuring of aTEF samples. Photolithography and ICP-RIE etching are performed for all samples, while resist removal and passivation are optional steps.} 
    \label{fig:atef-microstructuring}
\end{figure}

The fabrication of microstructured Si-PIN diodes was performed in \AP{a} cleanroom environment at MNF. A scheme with the individual fabrication steps is shown in fig. \ref{fig:atef-microstructuring}. A photoresist cover is \AP{spin-coated} on the diode and microstructured in a hexagonal pattern in a photolithography process. A silicon-etching process, the inductively-coupled plasma-reactive ion etching (ICP-RIE), follows. As a post-treatment, the photoresist \AP{is} removed from the diode surface, and an optional passivation layer \AP{is} applied. The respective steps are described in more detail in the following.
\paragraph{Photolithography}
The aim of the photolithography step is to coat the sample with a structured photoresist layer 
\AP{for etching} the desired structure in the silicon afterwards. The permanent epoxy negative photoresist Kayaku Microchem\,\footnote{Kayaku Advanced Materials, 200 Flanders Road, Westborough, MA 01581, USA} SU-8 3035\,\cite{Kayaku} was used as coating material. After being disposed on the sample, it was spun in a SPS europe\,\footnote{SPS Germany, Ferdinand-Braun-Straße 16, 85053 Ingolstadt, Germany} POLOS Spin 150i spin coater, first \AP{at} an acceleration of \SI{100}{rpm\per\second} to \AP{reach} \SI{1500}{rpm}, and after \SI{10}{s}, accelerated \AP{at} $\SI{300}{rpm\per\second}$ to \AP{reach} \SI{4000}{rpm}. The sample was then spun for \SI{90}{s} at this speed. A bake-out for up to \SI{15}{\minute} at \SI{95}{\degreeCelsius} on a hot plate followed for solvent removal. The resist-coated sample was then UV-illuminated in a EVG\,\footnote{EV Group Europe$\&$Asia/Pacific GmbH, DI Erich Thallner Strasse 1, A-4782 St. Florian am Inn, Austria} 620 NT Nano Imprint lithography system with a dose of \SI{250}{\milli\joule\per\square\centi\meter} through a Cr lithography mask, which was fabricated via electron-beam lithography with a Raith\,\footnote{Raith GmbH, Konrad-Adenauer-Allee 8, 44263 Dortmund, Germany} EBPG5150 system. The mask had a hexagonal structure with edge length $s = \SI{100}{\micro\meter}$ and wall thickness $d = \SI{10}{\micro\meter}$. After illumination, the sample was heated to \SI{65}{\degreeCelsius} for \SI{1}{\minute} and then to \SI{95}{\degreeCelsius} for \SI{10}{\minute} before developing the photomask in propylene glycol methyl ether acetate (PGMEA) for \SI{8}{\minute}. To stop the development process, the sample was then cleaned in isopropanol and dried with nitrogen \AP{afterwards}.\\
\begin{figure}
    \centering
    \includegraphics[width=0.3\linewidth]{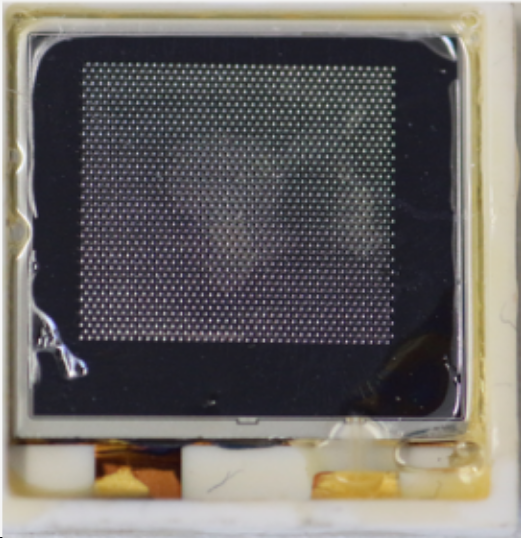}
    \caption{Picture of a $(1\times1)\,$cm$^2$ Hamamatsu S3590 photodiode after silicon etching.}
    \label{fig:diodephoto}
\end{figure}
The minimum hexagon-wall thickness $d$ achieved in the tests was limited by the ceramic packaging of the diodes, see fig. \ref{fig:diodephoto}. It overtopped the diode by \SI{0.7}{\milli\meter} \AP{around the edges}. Due to the resulting distance between \AP{the} lithography mask and \AP{the} diode, blurring effects occurred. \AP{The} achieved wall thicknesses \AP{typically} ranged from \SI{45}{\micro\meter} to \SI{60}{\micro\meter}, which is much larger than the target thickness of \SI{10}{\micro\meter}. Improvements were possible by switching to a diode type with ceramic packaging flushed with the silicon surface at a later stage, with which \deleted{the final} wall thicknesses between \SI{20}{\micro\meter} and \SI{30}{\micro\meter} were obtained. The etching process was not directional enough to achieve smaller thicknesses without partially destroying the walls by etching. \added{Although the etching results seemed more promising than in the first iteration, diodes of this different type showed worse performance already before etching (higher noise level, higher reverse current), and could not overcome the performance of the first-iteration diodes. Hence, the results presented in sec. \ref{sec:measurements} make use of diodes with standard packaging, despite the thicker channel walls. Only for the studies on the impact of surface passivation discussed in sec. \ref{sec:passivation}, diodes of the second iteration with thinner walls are used.}
\paragraph{Silicon etching}
\begin{figure}
    \centering
    \includegraphics[width=0.35\linewidth]{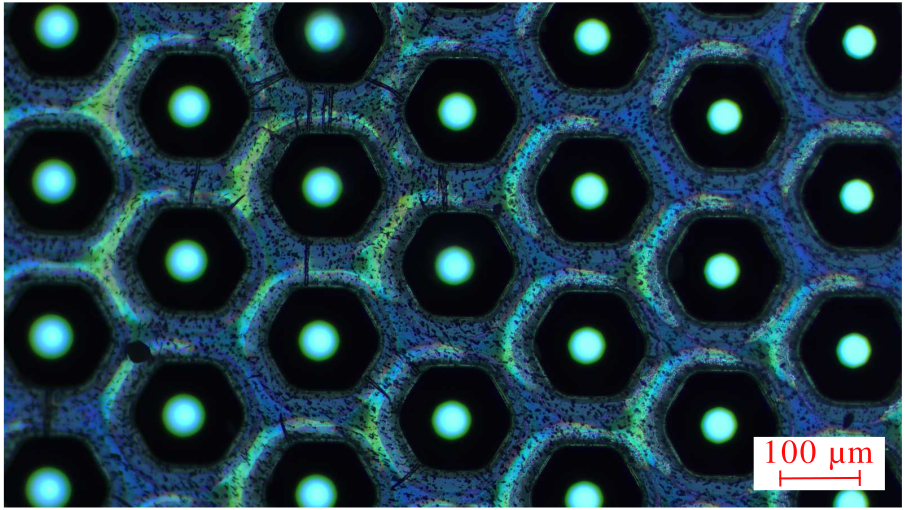}\\
  \includegraphics[width=0.35\linewidth]{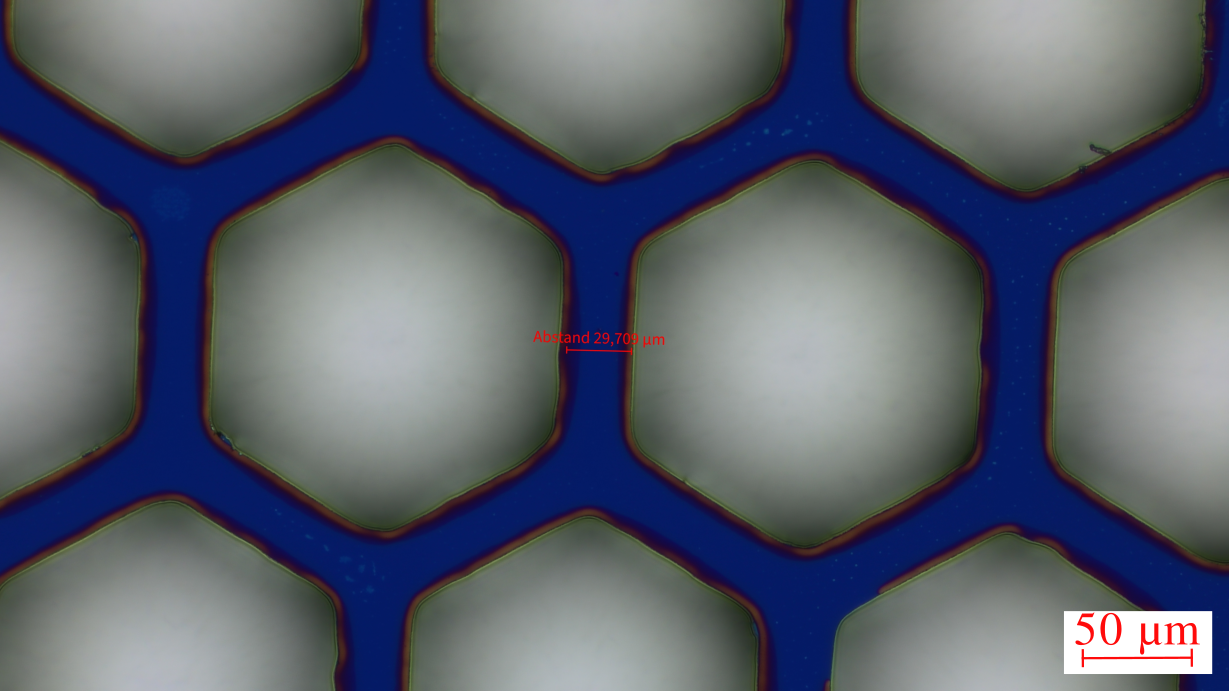}\\
    \includegraphics[width=0.35\linewidth]{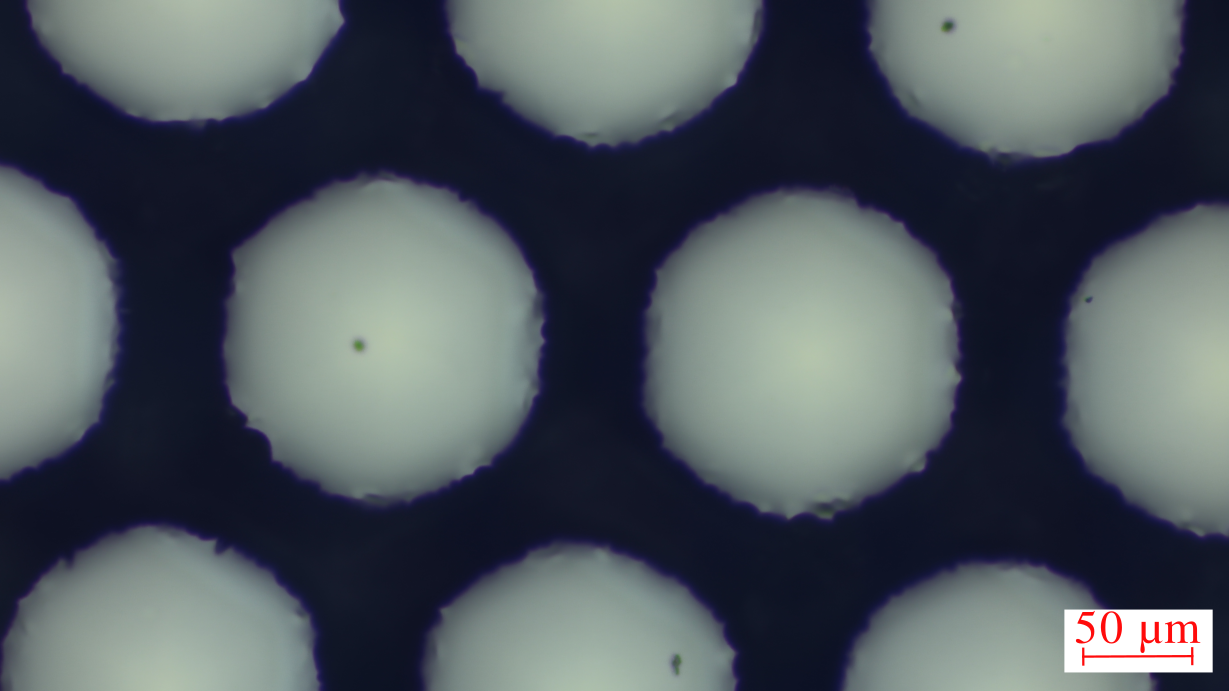}
\caption[caption]{Images of microstructured Si-PIN diodes fabricated at MNF via ICP-RIE. Top: Sample with $d =\SI{60}{\micro\meter}$ and photoresist on top. \added{With this sample, the measurements shown in sec. \ref{sec:measurements} were performed.} Middle: Sample with $d=\SI{30}{\micro\meter}$ \added{obtained using diodes with ceramic packaging flushed with the silicon surface}\replaced{. T}{, t}he photoresist was removed via treatment in DMSO. Bottom: Same sample with focus on the channel \replaced{floors}{grounds}.}
\label{fig:etchresult}
\end{figure}

For \AP{etching the silicon,} 
a highly directional etching process is required. The method of inductively-coupled plasma-reactive ion etching (ICP-RIE) \cite{cryo1,cryo2,cryo3,Maluf_2004} was applied using an Oxford\footnote{Oxford Instruments GmbH, Borsigstraße 15A, 65205 Wiesbaden, Germany} PlasmaPro 100 system. The process combines chemical and mechanical etching with the process gases \ce{SF_6} and \ce{O_2}. Ion bombardment of the silicon sample, placed on a 4-inch \ce{Si-SiO_2} carrier wafer, leads to the formation of \ce{SiF_4} and \ce{SiO_xF_y}.
\ce{SiF_4} is gaseous at room temperature and can be pumped out of the chamber, once formed. The second reaction, forming \ce{SiO_xF_y}, leads to surface passivation as the compound attaches to the silicon surface. Via this method, a highly anisotropic etching process is achieved.\\
The etching rate \textcolor{black}{and level of etch anisotropy are} strongly temperature dependent, and temperature stabilization is realized via a continuous flow of \ce{LN_2}-cooled helium gas. The thermal contact between \AP{the} sample and \AP{the} carrier wafer is achieved via Lesker\,\footnote{Kurt J. Lesker Company GmbH, Fritz-Schreiter-Str. 18, 01259 Dresden, Germany} Santovac 5 polyphenyl ether vacuum oil. Exemplary etching results are shown in fig. \ref{fig:etchresult}.
\section{Proof-of-principle measurements}
\label{sec:measurements}
\begin{figure}
    \centering
    \includegraphics[height = 0.4\textheight]{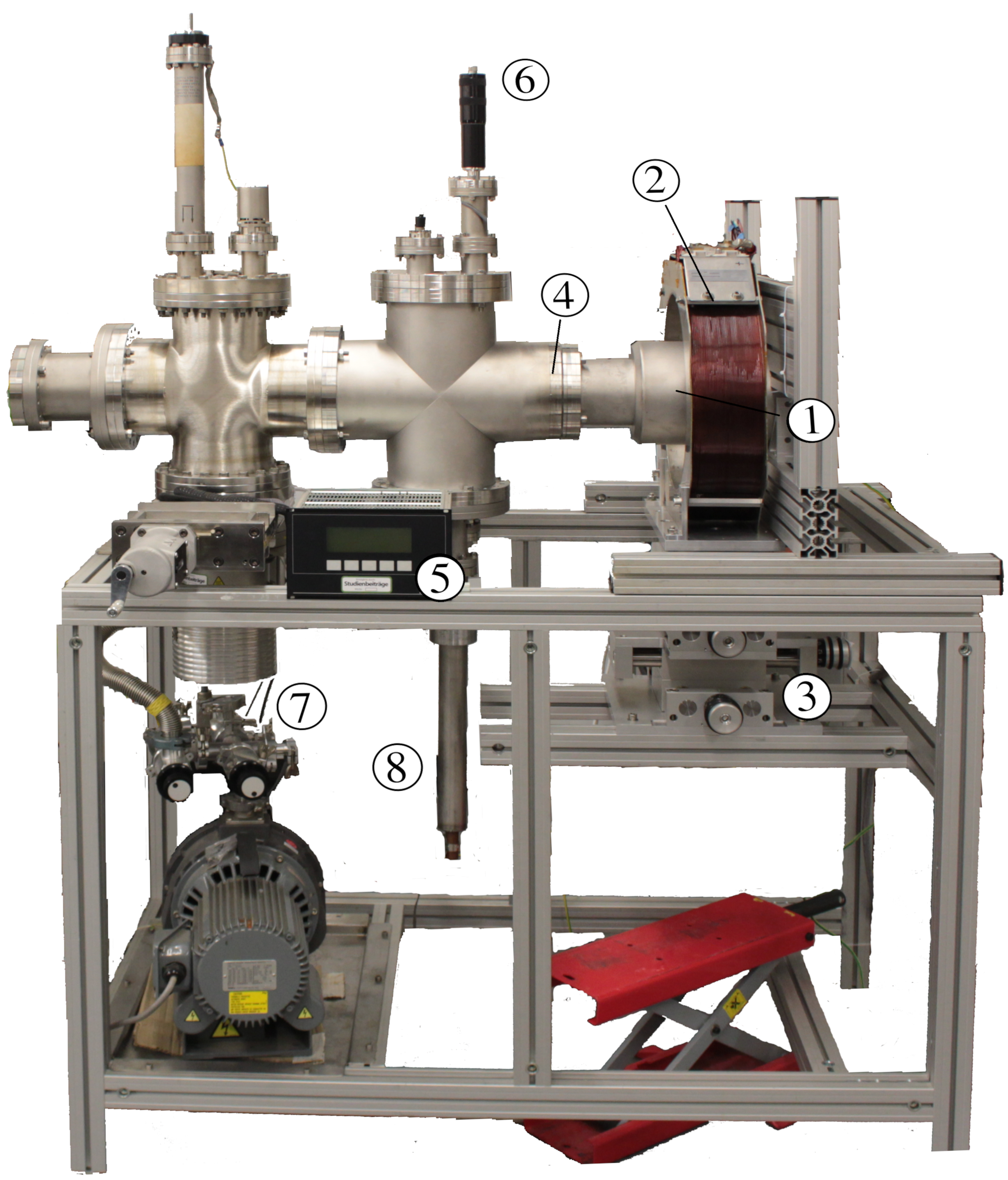}\\
        \includegraphics[height = 0.14\textheight]{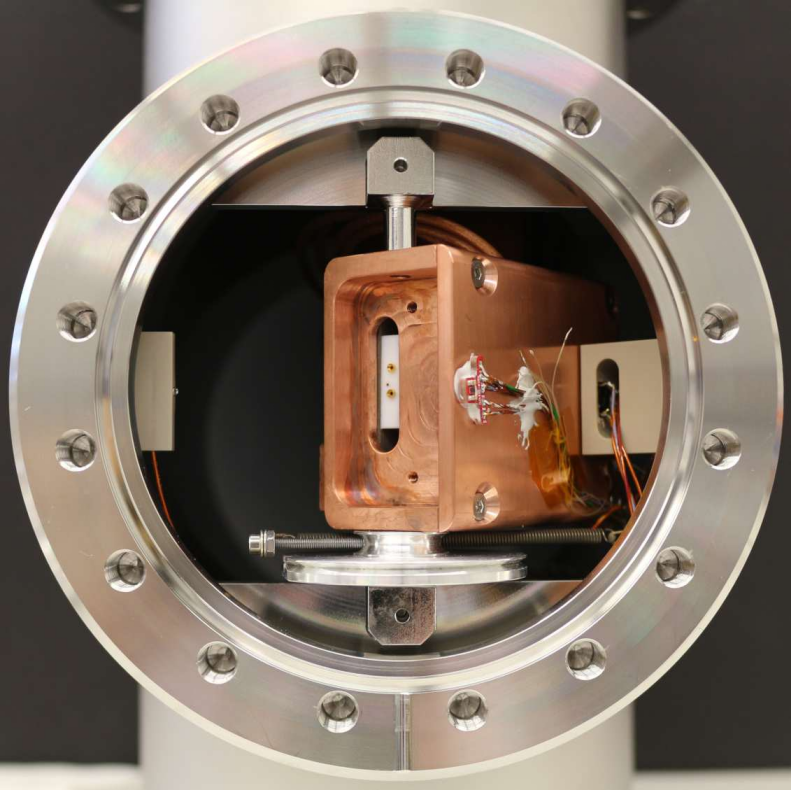}
        \includegraphics[height = 0.14\textheight]{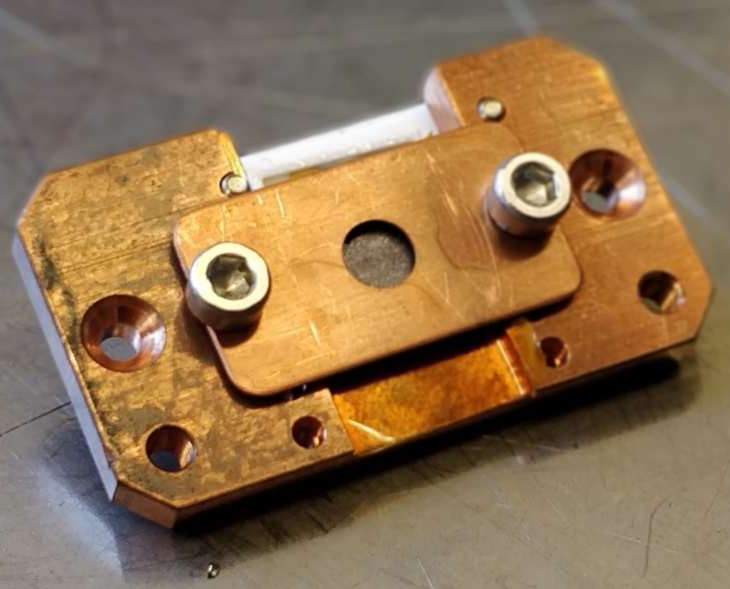}
    \caption{Top: Picture of the aTEF test setup. 1: Photoelectron source, 2: coil, 3: positioning tables, 4: detector and $1^{st}$ amplifier stage, 5: $2^{nd}$ amplifier stage, 6: linear feedthrough with Bowden cable, 7: Forepump and turbomolecular pump, 8: cold finger. Bottom: Rotatable copper case (left) containing the first amplifier stage. The detector holder (right) can be plugged onto the front of the case. Electrical contact is realized via plug contacts. The copper cover of the holder with a hole of \SI{5}{\milli\meter} diameter defines the detector region which is hit by the electron beam.}
    \label{fig:setup}
\end{figure}
\begin{figure}
    \centering
        \includegraphics[width=0.55\linewidth]{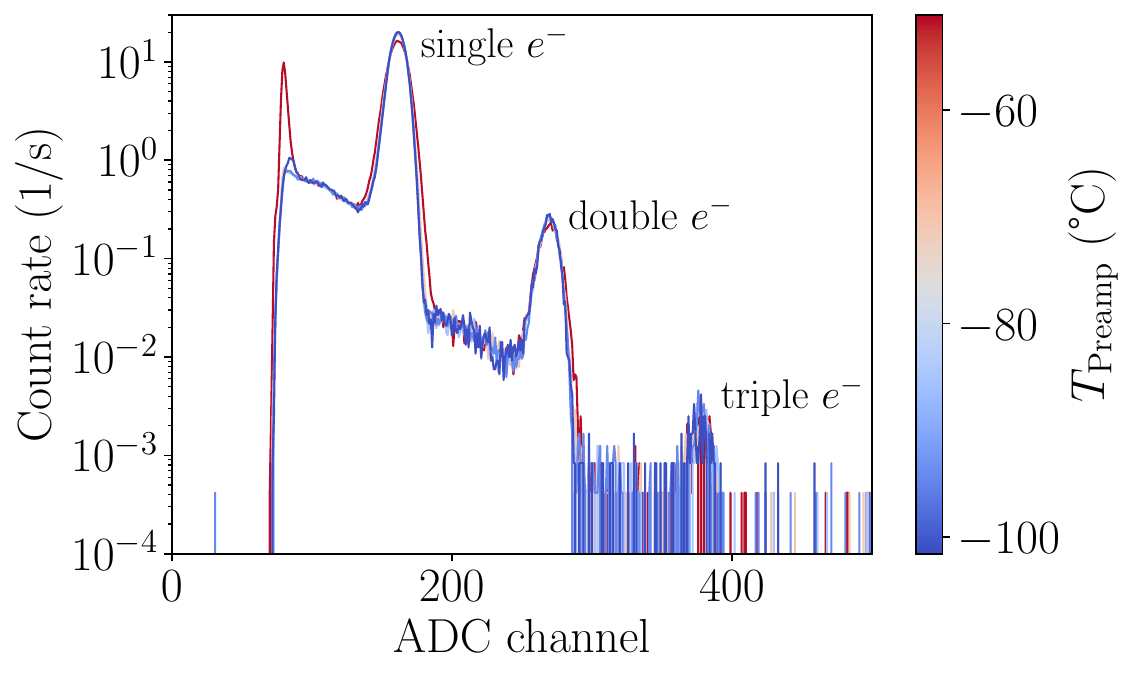}\\
    \includegraphics[width=0.55\linewidth]{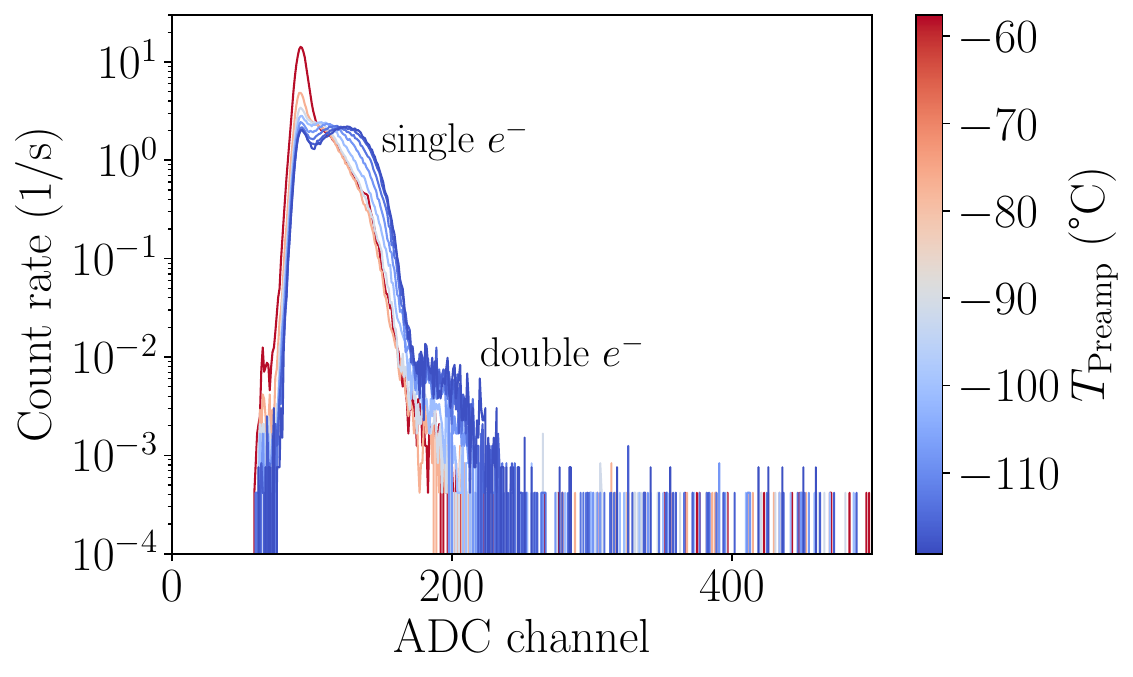}
    \caption{Spectra of \SI{20}{\kilo\electronvolt} electrons obtained during detector cool-down from room temperature to $\SI{-105}{\celsius}$, in the upper panel for a standard diode and in the lower panel for a microstructured one. The color scale indicates the preamplifier temperature. 
   The signal peak of the microstructured diode clearly shifts to higher ADC channels with decreasing temperature, indicating improvement of charge-collection efficiency.
   Nevertheless, the comparison of the two groups of spectra shows reduced charge-collection efficiency and energy resolution for the microstructured diode.}
    \label{fig:spectra-diodeD}
\end{figure}
\begin{figure}
    \centering
    \includegraphics[width=0.63\linewidth]{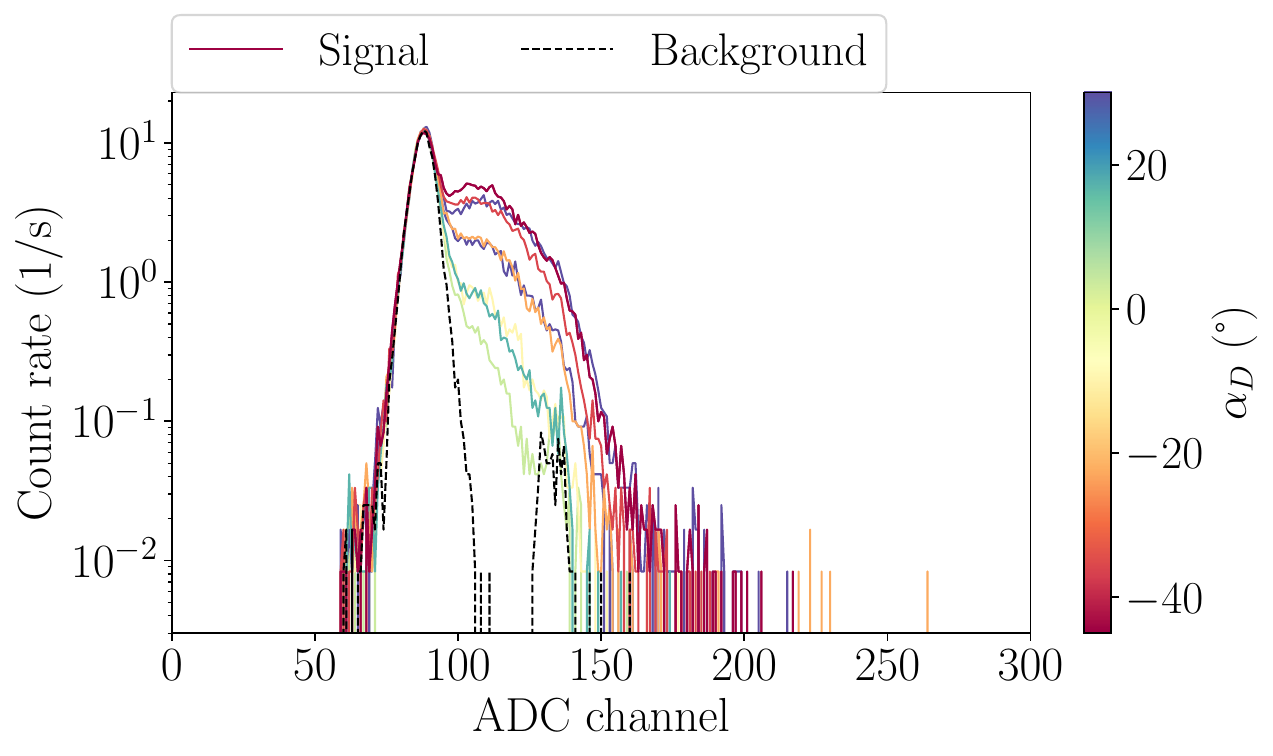}\\
   \hspace{-1.3cm} \includegraphics[width=0.5\linewidth]{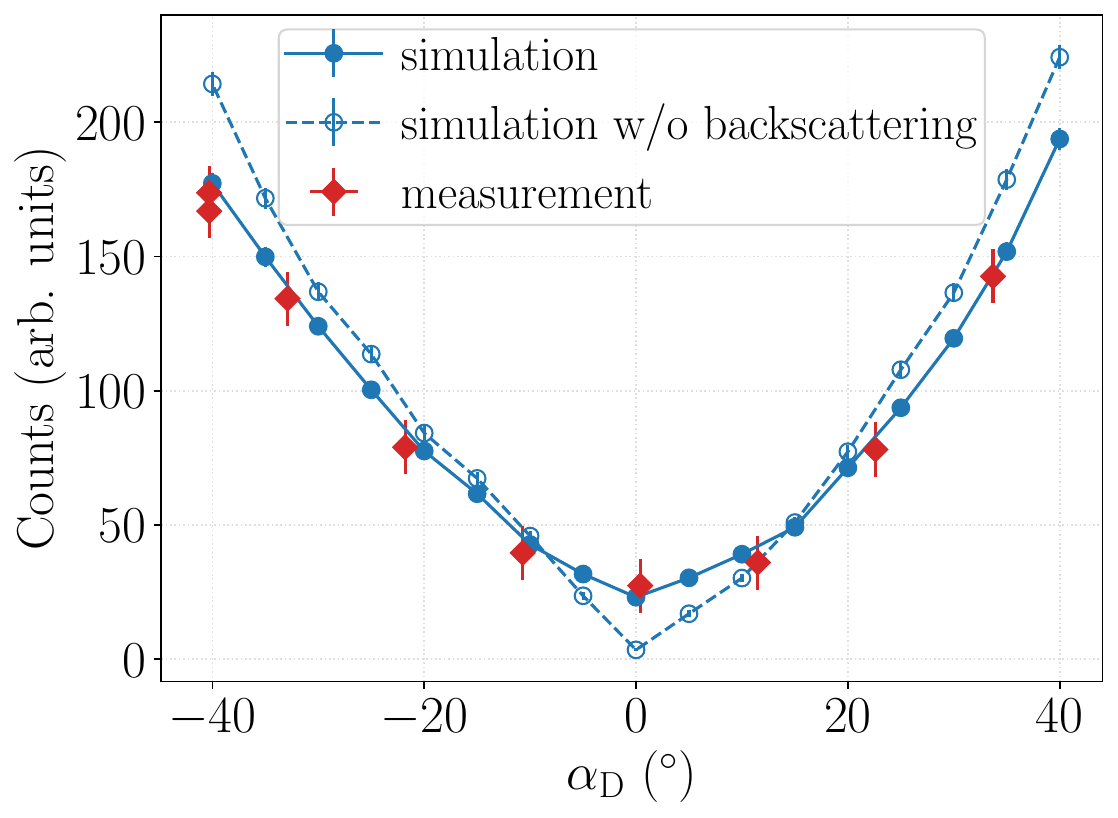}
    \caption{Top: Measured electron spectra for different incidence angles. Bottom: Comparison of measured angular-dependent count rates (red data points) with simulations (blue curves).}
    \label{fig:atef-winkel}
\end{figure}
\begin{figure}
     \centering
     \includegraphics[width=0.55\linewidth]{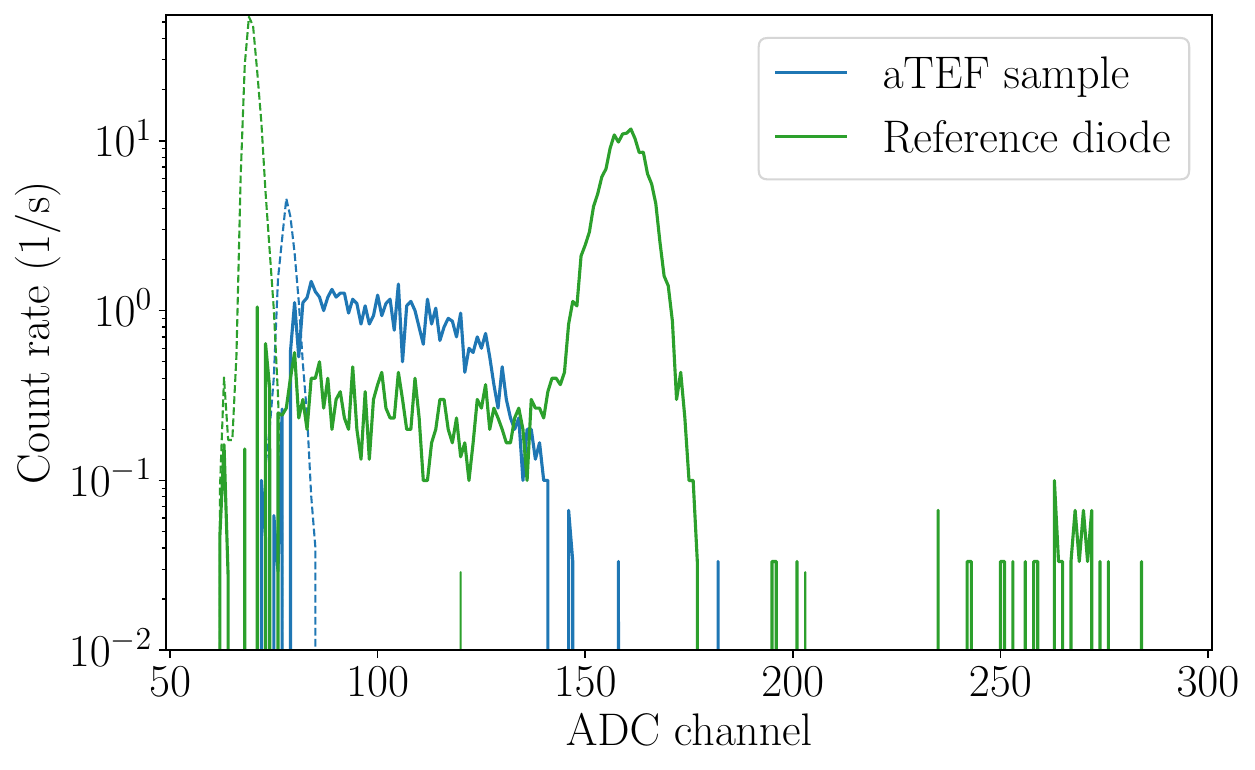}
          \begin{tabular}{lrr}
          \toprule
                                   & Reference diode & aTEF sample \\
          \midrule
       Cooling temperature ($^\circ$C)   & -50 & -107 \\
       Electron incident angle ($^\circ$) & 0 & 40 \\
       Bias voltage (V) & -60 & -18 \\
       Counts up to ADC channel 300 (cps)& $163.2\pm 9.4$ & $50.0\pm 2.4$\\
       \bottomrule
     \end{tabular}
         \caption{\added{Comparison of the measured spectrum of \SI{20}{\kilo\electronvolt} electrons obtained with the aTEF sample and a reference diode for determination of the electron-detection efficiency. The dashed lines show the noise spectrum measured without electron emission, which was subtracted from the total spectrum to obtain the signal spectra shown as solid lines. The table below the figure summarizes the measurement conditions in both measurements, which were conducted directly after another, with equal electron rate emitted from the electron source, as well as the sum of detected signal counts up to ADC channel 300. The uncertainty on the count rate includes the statistical uncertainty and a systematic uncertainty due to the background subtraction. }}
     \label{fig:absoluterate}
 \end{figure}

For proof-of-principle measurements regarding the expected angular-dependent detection efficiency of microstructured Si-PIN diodes, a dedicated test setup was built, shown in fig. \ref{fig:setup}.
Electrons are emitted by an in-house-fabricated angular-selective photoelectron source comparable to the one described in \cite{Behrens2017}, and guided towards the detector by a magnetic field of $\approx \SI{15}{\milli\tesla}$ strength produced by an
air coil. Typically, electrons with energies up to $\SI{20}{\kilo\electronvolt}$ and a narrow angular distribution along the magnetic-field lines are emitted with count rates of $\mathcal{O}(\SI{100}{cps})$ when being irradiated by a UV LED. The coil position and angle can be adjusted to steer the electron beam onto the detector. The detector is positioned in a rotatable holding structure visible in the lower left panel of the figure, which also contains the first stage of the capacitively coupled charge-sensitive amplifier readout. \added{The complete preamplifier readout is a customized setup which was originally developed for the Mainz Neutrino Experiment\,\cite{WEINHEIMER1992273}.} The structure can be tilted with the detector surface as pivot point via a linear feedthrough, which is connected to the holding structure via a Bowden cable. To exchange the detector, the flange between the electron source and the detector can be opened, and the source can be pulled back. The second amplifier stage is located outside vacuum. With a combination of a turbomolecular pump (Leybold TURBOVAC MAG W 400 iP) backed by an oil-free\deleted{ Leybold} forepump\deleted{\footnote{Leybold GmbH, Bonner Straße 498, 50968 Cologne, Germany}}\deleted{SC 15D}, a vacuum level of $<\SI{1e-7}{\milli\bar}$ is reached. The pressure readout is switched off during measurements to avoid disturbances. \replaced{Cooling of the detector and the first preamplifier stage}{Detector cooling} is possible via a cold finger supplied with \ce{LN_2}, the temperature at the detector can be varied by adjusting the penetration depth of the cold finger in the dewar. \added{The second preamplifier stage outside vacuum is at room temperature.} \\
The measured electron spectra of a standard Hamamatsu S3590 PIN diode as reference and a microstructured one, with their output pulses first amplified by a charge-sensitive preamplifier, then shaped by a shaping amplifier (time constant $\tau = \SI{0.5}{\micro\second}$) and read out by a \textcolor{black}{CAEN\footnote{CAEN S.p.A., Via Vetraia, 11, 55049 – Viareggio (LU), Italy} N957} multichannel analyser, are shown in fig. \ref{fig:spectra-diodeD}. The microstructured detector shows a broader charge response \AP{at the peak positions of the reference diode's charge spectrum, which indicates an inefficiency of its charge collection.}
It \AP{has} a worse energy resolution compared to the standard diode --- the differentiation between the 1-electron and the 2-electron pile-up peak is hardly possible. Cooling the detector to $\approx$ \SI{-100}{\celsius} reduces the noise level and improves the charge-collection efficiency. A reduced charge collection can occur due to a decreased charge carrier lifetime caused by trapping processes, which can be attributed to the large amount of defects in the microstructured diode induced by the etching process. Trapping effects are known to affect the pulse shape of silicon detectors \cite{Martini_1970}.
Electron spectra measured with the microstructured diode at different electron-incidence angles can be seen in fig. \ref{fig:atef-winkel} (top). The signal count rate after background subtraction as a function of \AP{the} detector tilt angle (and thereby, the electron incidence angle) is plotted in the lower panel of the figure. A clear increase in the \AP{net signal rate} with angle is observed. Also shown are two profiles from tracking simulations performed with the low-energy physics package \texttt{Penelope}\,\cite{Baro_1995,asai_penelope_2021} in \texttt{Geant4}\,\cite{Agostinelli_2003,Allison_2006,Allison_2016}.  \\
For the simulation, a simplified geometry with several hexagonal channels placed on a cylindrical bulk material with \SI{300}{\micro\meter} thickness was used. Channels with hexagon side length $s = \SI{80}{\micro\meter}$, $d = \SI{60}{\micro\meter}$ and a channel depth of $l = \SI{105}{\micro\meter}$ were assumed, in agreement with the real sample geometry determined with an optical microscope\,\cite{PhD-Gauda}\,\footnote{With the microscope, $d =  \SI{60}{\micro\meter}$ and a combined depth of channels and photoresist of \SI{130}{\micro\meter} were determined.}. A photoresist layer was simulated by an insensitive layer of \SI{25}{\micro\meter} thickness on top of the channel walls, with a density of \SI{1.2}{\gram\per\square\centi\meter} containing \SI{48.5}{\percent} \ce{H}, \SI{45.5}{\percent} \ce{O} and \SI{6.0}{\percent} \ce{C}. These material properties are based on the properties of Bisphenol A with the formula \ce{C_15H_16O_2}\,\cite{Lim_2007}, which is one of the main components of SU-8 resist. For each angle, 5000 electrons were simulated. When backscattering was not taken into account, each electron hitting the detector first in the sensitive channel area was tallied. In the case with backscattering included, each electron depositing at least \SI{12}{\kilo\electronvolt} in the sensitive volume was counted.
 The simulated counts were scaled by a common factor to match the measured counts, as the absolute detection efficiency of the sample is hard to determine within \replaced{this}{the} measurement.\\ 
 The measured profile matches well with the simulation when taking backscattering effects into account\footnote{It should be noted that several aspects are not accounted for in the simulation, such as the imperfect surface and the surface roughness of the etched channels, or changes of channel geometry along the depth profile caused by the nanofabrication process. Furthermore, the detection threshold of \SI{12}{\kilo\electronvolt} is an estimate based on experience with the samples, but was not verified quantitatively.}. \\
\added{To estimate the aTEF's detection efficiency relative to a standard diode, the measured count rates of the two detectors were compared in a dedicated measurement. A reduced electron rate was chosen to lower the impact of pile-up effects. The two taken spectra are shown in fig. \ref{fig:absoluterate}. The count rate measured with the aTEF sample at \SI{40}{\degree} incident angle is $(30.7\pm 2.3)\,\%$ of that of the reference diode. The tested sample has a hexagon wall thickness of \SI{60}{\micro\meter}. The channel tops are covered with photoresist and are thus insensitive, and the sensitive holes make only $(43\pm 2)\,\%$ of the total aTEF area. Therefore, $(43\pm 2)\,\%$ of the count rate measured with the reference diode are expected to be measured with the aTEF sample, if the efficiency is identical. The simulation predicts a detection of \SI{44}{\percent} of the incident electrons at \SI{40}{\degree} electron-incident angle, when taking the angular-dependent backscattering into account. Due to the lower rate observed in the measurement, it is concluded that the detection efficiency of the channels is approximately $(0.307\pm 0.023)/(0.43 \pm 0.02)=(0.71 \pm 0.06) = (71 \pm 6)\,\%$.}

\section{Resist removal and surface passivation}
\label{sec:passivation}
\begin{figure}
    \centering
    \includegraphics[height = 0.3\textheight]{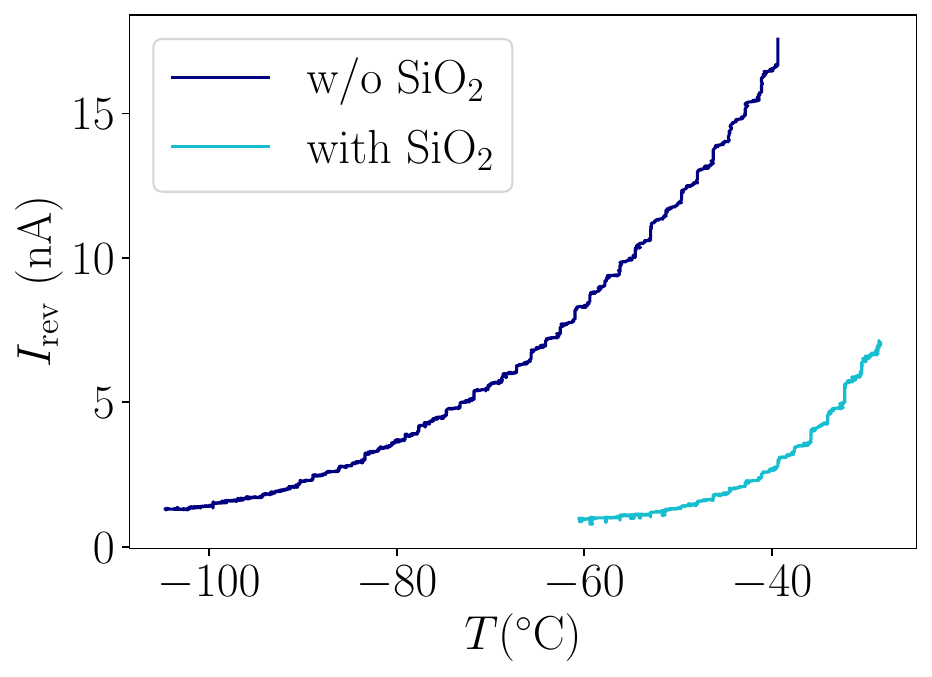}\\
        \includegraphics[height = 0.3\textheight]{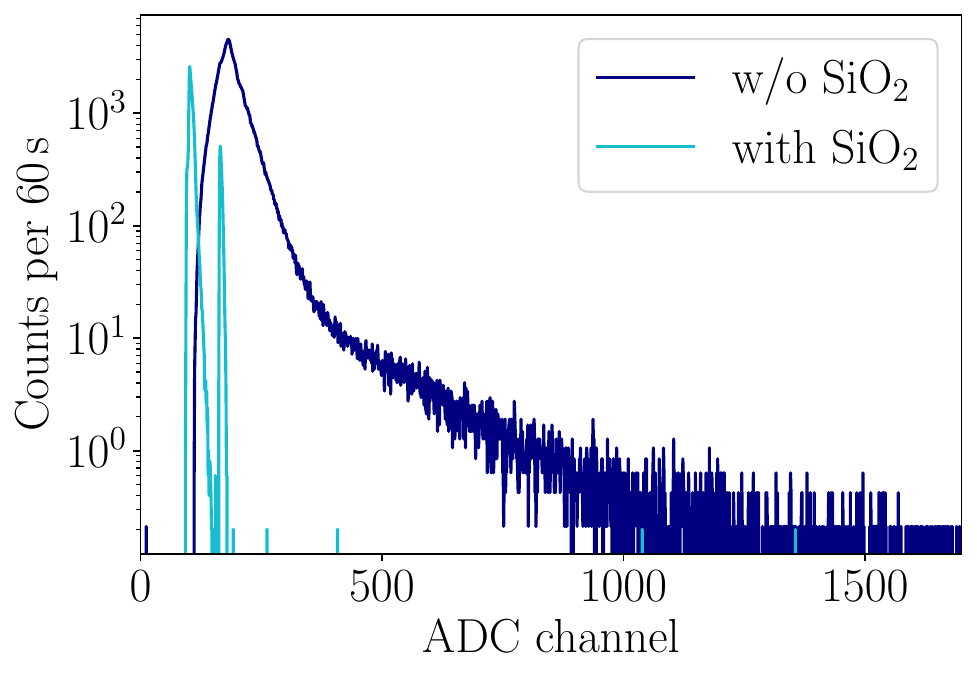}
    \caption{Effect of surface passivation with \ce{SiO_2} of $\SI{20}{\nano\meter}$ thickness. Top: Reverse current as a function of sample temperature before and after passivation. Bottom: Noise spectra at $\SI{-40}{\degreeCelsius}$ measured over \SI{60}{\second} with the same microstructured diode before and after passivation.}
    \label{fig:coating}
\end{figure}
The etching process can lead to a high number of crystalline defects, \AP{such as} vacancies or dangling bonds on the surface. These defects affect the energy states in the band gap, and lead to undesired dark current due to the formation of metallic conduction channels\,\cite{Mayet_2018,Tao_2003}. Also, charging effects can occur. \AP{Termination} of dangling bonds is possible \AP{through} surface passivation.\\
Before passivation, \AP{the remaining photoresist on the channel tops was removed}. It was realized by heating the diode in Dimethyl sulfoxide (DMSO) to up to \SI{80}{\celsius} for several hours \AP{that} DMSO soaks the SU-8 resist and allows its detachment, while the diode stays intact.\\
Common passivation procedures include hydrogenation in diluted hydrofluoric acid solution\,\cite{Burrows_1988}, a thermal oxidation at around \SI{800}{\degreeCelsius}\,\cite{Schultz_2008}, plasma-enhanced chemical vapour deposition (PECVD)\,\cite{kim_2011}, or the deposition of thin dielectric layers like \ce{SiO_2} or \ce{Al_2O_3} via atomic layer deposition (ALD)\,\cite{Pasanen_2017}. All these approaches are reported to suppress dark current by three to four orders of magnitude\,\cite{Mayet_2018}. Crucial for \AP{selecting} a passivation procedure for \AP{the} aTEF diodes was that the diode should not be destroyed during passivation, and that the process should be \replaced{realizable}{realisable} on-site in Münster. In addition, it was desirable that the electrodes required for contacting the diode should not be affected. A thermal oxidation was not possible because the diodes should not be exposed to temperatures above \SI{80}{\degreeCelsius}\,\cite{Hamamatsu}, and a hydrogenation process was not feasible in the given infrastructure. A PECVD process would be an option, but it \AP{has not been tested} so far. Passivation tests with \ce{Al_2O_3} deposited via ALD were performed, but yielded irreproducible results. Better results were obtained via thermal evaporation of \ce{SiO_2} in a Physical Vapour Deposition (PVD)\,\cite{Mattox_2014} setup: \ce{SiO_2} was heated in a sample holder under vacuum conditions of $\mathcal{O}(\SI{1e-6}{\milli\bar})$ by \AP{the} application of high currents up to \SI{300}{\ampere}. \ce{SiO_2} has a melting point of around \SI{1700}{\degreeCelsius}\,\cite{weast_handbook_1984} and is thus harder to evaporate than other typical materials such as Ag, Au, or Cr. Nevertheless, evaporation was possible, and films with \qtyrange[range-units = bracket]{5}{50}{\nano\meter} thickness\footnote{The passivation layer is so thin that electrons can still pass through and enter the active volume.} were deposited, using a deposition rate of $\mathcal{O}{(\SI{1}{\angstrom\per\second})}$. During the first deposition tests, the final film thickness was cross-checked with a high-accuracy weighing machine. The sample was water-cooled during evaporation to reduce its temperature. \\
A measurement of the reverse current\footnote{\textcolor{black}{Reverse current here means an undesirable dark current in the reverse direction of the diode if a bias voltage is applied. The nominal reverse current of Hamamatsu S3590-09 diodes is \qtyrange[range-units=bracket]{2}{6}{\nano\ampere} at \SI{70}{\volt}\,\cite{Hamamatsu}.}} $I_\text{rev}$ as a function of temperature during cooling of a microstructured diode in the aTEF test setup before and after coating with \SI{20}{\nano\meter} \ce{SiO_2} is shown in fig. \ref{fig:coating} (upper panel). While $I_\text{rev} <\SI{2}{\nano\ampere}$ is reached at \SI{-92}{\degreeCelsius} sample temperature without \AP{passivation}, this reverse-current level is obtained already at \SI{-42}{\degreeCelsius} with the same diode after \AP{the} \ce{SiO_2} coating. The reduction of $I_\text{rev}$ is accompanied by a strong noise reduction, especially of the shot-noise contribution, as visible in the lower panel of the figure, comparing the noise spectra of the diode at \SI{-40}{\degreeCelsius} before and after coating. The impact of different passivation thicknesses was investigated and thicknesses of \SI{20}{\nano\meter} and \SI{50}{\nano\meter} were found to yield comparable performance improvement, with slightly better results for a thickness of \SI{50}{\nano\meter}.\\
Despite these performance improvements, the additional treatment steps apparently damaged the prototypes. \AP{Notably}, all of them \AP{exhibited} a reduced signal rate after the removal of the photoresist, and the passivation did not lead to an improvement in detection efficiency. In addition, the energy resolution \AP{deteriorated further} after passivation. It is expected that these degradations could be avoided by improving the various nanofabrication steps to achieve smooth channels with \AP{fewer} vacancies and a homogeneous, impurity-free passivation layer. Also, the process-related heating to up to $\SI{130}{\celsius}$ during passivation could be \AP{a} cause of the performance degradation. 
\AP{To address these issues,} the fabrication of additional aTEF samples was outsourced to the Fraunhofer IZM\footnote{Fraunhofer-Institut für Zuverlässigkeit und Mikrointegration IZM, Gustav-Meyer-Allee 25, 13355 Berlin, Germany}. Results obtained with etched diodes produced by IZM will be published elsewhere. 
\section{Summary and outlook}
\label{sec:summary}
We have introduced the concept of a low-energy electron detector with specific angular response (aTEF) based on a microstructured Si-PIN diode. It is able to suppress electrons with low incident angles relative to the detector's surface normal\deleted{or to the magnetic field axis in case of magnetically guided electrons, and thus discriminate between different cyclotron radii}.
\replaced{Applying microstructuring by ICP-RIE at the Münster Nanofabrication Facility to}{Using microstructured} commercial $(1\times 1)\,$cm$^2$ Hamamatsu Si-PIN diodes\deleted{ as aTEFs}, we successfully demonstrated \replaced{the aTEF}{this} principle in a test setup using an angular-selective photoelectron source. \\
Although the microstructuring procedure caused a significant \added{increase in leakage current and a} reduction in \added{charge collection and} energy resolution, \replaced{a detection efficiency of the microstructured channel walls of $(71\pm 6)\,\%$ at \SI{40}{\degree} incident angle to the detector surface normal for \SI{20}{\kilo\electronvolt} electrons was achieved. The leakage current was reduced  by cooling the microstructured detector to down to \SI{-100}{\celsius}, the overall performance might be further improved by modified microstructuring and surface-passivation methods.}{the ansatz might be improved by modified microstructuring and passivation methods.}\\
 As an application, such detectors could serve to reduce electron background at the KATRIN experiment, which has a significantly different angular distribution than the signal electrons. \added{In the multiple Tesla-strong magnetic field at the KATRIN detector, the electrons gyrate around magnetic-field lines, thus the different electron angles correspond to different cyclotron radii being discriminated by the channel geometry of the aTEF.}
Assuming the background distribution shown in fig. \ref{fig:background_at_KATRIN}, the aTEF geometry for optimal signal-to-background ratio at KATRIN was determined to have a hexagon side length $s = \SI{80}{\micro\meter}$, a channel depth of $l = \SI{250}{\micro\meter}$ and a wall thickness of $d = \SI{10}{\micro\meter}$\,\footnote{The wall thickness and depth are a compromise between a good open-area ratio, good angular selectivity and a sufficient mechanical stability of the detector.}. \added{The parameters were obtained using a dedicated Monte-Carlo electron-tracking simulation which is described in detail in \cite{Gauda_2022} and uses the specific experimental parameters of the KATRIN setup.} With such a geometry, a background reduction by a factor between 3 and 5 would be possible, depending on the fraction of autoionizing states contributing to the background, otherwise produced by Rydberg atoms\added{, while retaining around \SI{80}{\percent} of signal efficiency, if the detection efficiency of the channel walls can be enhanced to the one of a commcercial diode. Estimating the impact on the squared neutrino-mass uncertainty $\delta m_\nu^2$, which is the key parameter at KATRIN, via \,\cite{KATRINDesignReport}
\begin{equation}
    \delta m_\nu^2 = \left( \frac{16}{27}\right)^2\cdot R_\text{s}^{-2/3} \cdot R_\text{b}^{1/6}\cdot t^{-1/2}
\end{equation}
for a signal rate $R_\text{s}$, background rate $R_\text{b}$ and measurement time $t$ delivers an expected improvement of the sensitivity on the squared neutrino mass by $\mathcal{O}(\SI{10}{\percent})$.} \\The aTEF prototypes used in this work \added{are not suited to allow this improvement in neutrino-mass sensitivity, since they} all had thicker and shorter channels, \added{coming along with an absolute detection efficiency of about \SI{31}{\percent} when considering the channel tops covered with photoresist, even at high incident angles.} \added{In addition, the tested samples} were much smaller than the actual \SI{9}{cm} \added{diameter} KATRIN detector, and suffered from inhomogeneities and an overall rough surface after microstructuring. The challenges in nanofabrication are expected to be overcome by the fabrication at IZM. The bigger challenge at KATRIN is the reverse $n^+np^+$ doping-layer structure of its detector. In contrast to the $p^+nn^+$ diodes studied here, the $n^+$ layer needs to be structured, and the depletion zone would spread from the bulk towards the channels. Consequently, the channel \replaced{floors}{grounds} could be sensitive as well and would require an additional passivation layer. Respective performance tests with a reverse doping-layer structure are underway.\\
\added{Anyway, the first phase of the KATRIN experiment, measuring for \SI{1000}{days} the tritium $\upbeta$-decay spectrum near its endpoint, has been finished by the end of 2025, and the second phase of KATRIN with the TRISTAN detector \cite{Houdy_2020} investigating the tritium $\upbeta$-decay spectrum nearly over the full energy range to search for signatures of keV sterile neutrinos is starting soon\,\cite{KATRIN_input_ESPP2024}. While an excellent detection efficiency is essential for a low-event experiment like KATRIN, a lower efficiency might be acceptable when making use of the angular-dependent detection efficiency for a characterization of the angular distributions of signal and background electrons, or in a design with large count rates, as is the case exemplarily in the upcoming TRISTAN phase of KATRIN\,\cite{Houdy_2020}.}

\appendix
\section{Derivation of the electric potential in a PIN diode microstructured from the $p^+$ side\label{appendix}}
The general solution of the Poisson equation is
\begin{equation}\label{eq:efield}
	E(x) = \frac{1}{\varepsilon} \int_{-\infty}^{x} \rho(x')\,\mathrm{d}x'
\end{equation}
with electric field $E(x)$ at depth $x$ in the PIN diode.\\
Let the microstructure have a depth $x_\mathrm{aTEF}$. For $w<x_\mathrm{aTEF}$, the electric field is
\begin{equation}\label{eq:tef_efield_atef_p0}
	\begin{split}
		E(x) 	&= \frac{1}{\varepsilon} (-e N_A x_p + e N_D x) \\
		&= \frac{eN_D}{\varepsilon} (x - w)\,.
	\end{split}
\end{equation}
With increasing $U_\text{bias}$, $w > x_\mathrm{aTEF}$ is possible. Then, the electric field in the region $0<x<x_\mathrm{aTEF}$ becomes
\begin{equation}\label{eq:tef_efield_atef_p1}
	\begin{split}
		E(x) 	&= \frac{eN_D}{\varepsilon} \left(x - x_\mathrm{aTEF} - \frac{w-x_\mathrm{aTEF}}{\delta} \right) \\
		&=: \frac{eN_D}{\varepsilon} (x-w_1)
	\end{split}
\end{equation}
with $w_1 = x_\mathrm{aTEF} + \delta^{-1} \cdot (w - x_\mathrm{aTEF})$ and the reduction factor $\delta = 1-\mathrm{OAR}$ depending on the open-area-ratio OAR of the microstructure.\\
For $x>x_\mathrm{aTEF}$, the electric field is given by
\begin{equation}\label{eq:tef_efield_atef_p2}
	\begin{split}
		E(x) 	&= \frac{1}{\varepsilon} \left(-e N_A x_p + e N_D x_\mathrm{aTEF} + e N_D \frac{(x-x_\mathrm{aTEF})}{\delta}\right) \\
		&= \frac{eN_D}{\varepsilon} \left(\left(-x_\mathrm{aTEF} -\frac{w-x_\mathrm{aTEF}}{\delta}\right) + x_\mathrm{aTEF} + \frac{x-x_\mathrm{aTEF}}{\delta}\right) \\
		&= \frac{eN_D}{\varepsilon} \frac{x-w}{\delta}\,.
	\end{split}
\end{equation}
For the electric potential, it follows
\begin{equation}
	\begin{split}
		U(x) &= \int_{-\infty}^x E(x') \,\mathrm{d}x' \\
		&=	\frac{eN_D}{\varepsilon}
		\begin{cases} 
			\frac{x^2}{2} - xw &\mathrm{for}\quad U < U_\mathrm{aTEF}\,,\\\\
			x_\mathrm{aTEF} \left( \frac{x_\mathrm{aTEF}}{2} - w_1 \right) - \frac{x_\mathrm{aTEF}}{\delta} \left( \frac{x_\mathrm{aTEF}}{2} - w \right) + \frac{x}{\delta} \left( \frac{x}{2} - w \right) &\mathrm{for}\quad U > U_\mathrm{aTEF}\,,
		\end{cases}
	\end{split}
\end{equation}
\section*{Acknowledgments}
We acknowledge the support of Ministry for Education and Research \replaced{BMFTR}{BMBF} (contract number 05A23PMA) and Deutsche Forschungsgemeinschaft DFG (Research Training Group GRK 2149) in Germany. 
We further acknowledge the Münster Nanofabrication Facility (MNF) for their support during the fabrication of silicon microstructure samples. We kindly thank Joscha Lauer (KIT Karlsruhe) for making his Geant4 code accessible, on which the simulations were based. We furthermore thank Norbert Wermes (University of Bonn) for his valuable input regarding silicon-detector etching and applications at the ATLAS detector\textcolor{black}{. We kindly thank Alan Poon (Berkeley Lab) for his helpful comments during reviewing the paper draft}.
\section*{Declarations}
\paragraph{Funding}
This work was supported by the German Ministry for Education and Research \replaced{BMFTR}{BMBF} (05A20PMA, 05A23PMA) and Deutsche Forschungs\-gemeinschaft DFG (Research Training Group GRK 2149).
\paragraph{Conflict of interest/Competing interests}
The authors have no relevant financial or non-financial interests to disclose.
\paragraph{Availability of data and materials}
The datasets generated during and/or analyzed during the current study are available from the corresponding author on reasonable request.
\paragraph{Code availability}
Not available
\noindent
\bibliography{bibliography.bib}

\end{document}